\documentclass[aps,twocolumn,amsmath,amsfonts,showpacs]{revtex4-1}

\usepackage{epsf}

\usepackage{graphicx}
\usepackage{amsfonts}
\usepackage{amsmath}
\usepackage{amssymb}
\usepackage{graphicx}
\usepackage{multirow}

\def\bbbc{{\mathchoice {\setbox0=\hbox{$\displaystyle\rm C$}\hbox{\hbox
to0pt{\kern0.4\wd0\vrule height0.9\ht0\hss}\box0}}
{\setbox0=\hbox{$\textstyle\rm C$}\hbox{\hbox
to0pt{\kern0.4\wd0\vrule height0.9\ht0\hss}\box0}}
{\setbox0=\hbox{$\scriptstyle\rm C$}\hbox{\hbox
to0pt{\kern0.4\wd0\vrule height0.9\ht0\hss}\box0}}
{\setbox0=\hbox{$\scriptscriptstyle\rm C$}\hbox{\hbox
to0pt{\kern0.4\wd0\vrule height0.9\ht0\hss}\box0}}}}

\begin{document}
\title{Structure of the number projected BCS wave function}

\author{ J.~Dukelsky$^{1}$, S. Pittel$^{2}$, and C. Esebbag$^{3}$}

\affiliation{
$^{1}$ Instituto de Estructura de la Materia, CSIC, Serrano 123,
28006 Madrid, Spain\\
$^{2}$ Bartol Research Institute and Department of Physics and
Astronomy, University of Delaware, Newark, DE 19716 USA \\
$^3$ Departamento de F\'isica y Matem\'aticas, Universidad de Alcal\'a, 28871
Alcal\'a de Henares, Spain}

\begin{abstract}
We study the structure of the number projected BCS (PBCS) wave function in the particle-hole basis, displaying its similarities with coupled clusters theory (CCT). The analysis of PBCS together with several modifications suggested by the CCT wave function is carried out for the exactly solvable Richardson model involving a pure pairing hamiltonian acting in a space of equally-spaced doubly-degenerate levels. We point out the limitations of PBCS to describe the non-superconducting regime and suggest possible avenues for improvement.
\end{abstract}


\author{}

\maketitle

\section{Introduction}

Pairing correlations are widespread over several areas of quantum-many body physics, ranging from condensed matter to quantum chemistry to cold atomic gases to atomic nuclei. The standard formalism for the description of these correlations is through the use of the Bardeen Cooper Schrieffer (BCS) approximation \cite{BCS}, whereby the correlations are described by means of a coherent state of collective pairs that explicitly breaks the conservation of particle number. This approach was extremely successful in the description of condensed matter superconducting systems with a macroscopic number of interacting electrons where fluctuations in the particle number are negligible. For systems with a fairly small number of particles, e.g. atomic nuclei or superconducting grains, it is important to restore particle number, through the use of the number projected BCS (PBCS) approximation \cite{PBCS}. PBCS has been an especially useful method to describe superconducting nuclei with few nucleons in the valence space \cite{Schmid}. However, when applied to ultrasmall superconducting grains \cite{Delft} it was revealed to be unable to describe the disappearance of superconductivity for small enough grains \cite{Grain}.  As a consequence of these studies, the exact solution of the reduced BCS Hamiltonian found by Richardson in the sixties \cite{Richardson} was rediscovered \cite{Delft2} and exploited as a powerful benchmark for many-body approximations.
Likewise, more general pairing Hamiltonians are exactly solvable if they can be expressed as a linear combination of the set of integrals that define the Richardson-Gaudin models \cite{RG1}. This exact solvability has enabled the test of approximate methods of treating pairing for a wide variety of systems, like small superconducting grains \cite{Sierra2}, realistic atomic nuclei \cite{Bang, Hirsch, Dussel, Romb}, and more recently in quantum chemistry \cite{Ayers}. Such tests have illustrated that for a large enough number of active orbits and weak pairing, the PBCS approximation misses important pairing correlations, making its use in large-scale energy density functional treatments of finite nuclei suspect. This has led to a multitude of efforts to develop improved approximate treatments of pairing correlations. This includes, {\it e.g.}, the use of RPA methods \cite{Hirsch} and the use of coupled cluster methods \cite{Scu1, Scu2}.

In this work we study the accuracy of the PBCS approximation in the non-superconducting regime and propose alternative methods based on a generalization of the PBCS wave function for an improved approximate treatment of pairing correlations. The method starts with the PBCS wave function, which is then expanded in the particle-hole (ph) basis. In the PBCS approximation, each term in the series expansion is defined by the expansion coefficients of a single collective pair and furthermore the contribution of each term is prescribed. The similarity of the PBCS wave function with the RPA in the quasiboson approximation and with the CCT of doubles with Seniority zero suggests several possible improvements that we explore in this work.

The structure of the paper is as follows. In Section II, we describe the PBCS approximation and several improvements based on the CCT. In Section III we introduce the exactly solvable pairing Hamiltonian that we use to carry out comparative tests of the various approximations and then in Section IV describe the results of this comparison and draw some conclusions.
In Section V we summarize the main results of the work and outline some issues for future consideration.

\section{The PBCS wave function in the particle-hole basis}

Let us consider a set of $N$ particle pairs moving in a space of $L$ doubly-degenerate single-particle states $i, \bar{i}$ and
denote the single-particle creation and annihilation operators associated with these states as $c^{\dagger}_i$, $c^{\dagger}_{\bar{i}}$ and $c_i$, $c_{\bar{i}}$, respectively. Furthermore, we denote the operators that create and annihilate a pair of particles in doubly-degenerate time-reversed states $i,{\bar{i}}$ as

\begin{eqnarray}
P_{i}^{\dagger}= c_{i}^{\dagger} c_{\bar{i}}^{\dagger} ~,\\
P_{i}= \left[ P_{i}^{\dagger} \right]^{\dagger} = c_{\bar{i}} c_i ~,
\end{eqnarray}
which satisfy the usual $SU(2)$ commutation relations
\begin{equation}
\left[ P_{i},P_{j}^{\dagger }\right] =\delta _{ij}\left( 1-N_{i}\right)~,
\qquad \left[ N_{i},P_{j}^{\dagger }\right] =2\delta _{ij}P_{j}^{\dagger }~,
\label{commu}
\end{equation}
where $N_{i}=c_{i}^{\dagger }c_{i}+c_{\bar{i}}^{\dagger }c_{\bar{i}}~$.

For simplicity we will restrict all our derivations to the seniority zero ($v=0$) subspace. However, the extension to broken pairs is straightforward and does not add any qualitative difference to our main conclusions.
We start from the BCS wave function, which in the ph basis can be written as (see, for example, Ref. \cite{RS})
\begin{equation}
\left| \text{BCS}\right\rangle =\frac{1}{\sqrt{\textsf{N}_{BCS}}} \exp{\sum_{j=M+1}^{L} \frac{v_j}{u_j} P^{\dagger}_j}   \times \exp{\sum_{i=1}^{M} \frac{u_i}{v_i} P_i}  \left|\text{ HF}\right\rangle ~,
\label{BCS}
\end{equation}
where $M=N/2$ is the number of pairs, $\textsf{N}_{BCS}$ is a normalization constant, and the HF state corresponds to having the lowest $M$ orbits filled. Clearly, the BCS wave function does not preserve particle number. Restoring particle number in this ph representation is just a matter of selecting in the expansion of both exponential the same number of particle creation pairs as hole destruction pairs. The final result is:

\begin{equation}
\left| \text{PBCS}\right\rangle =\frac{1}{\sqrt{\textsf{N}_{PBCS}}} \sum_{l=0}^{L} \frac{1}{(l!)^2} \left[ \sum_{i=1,j=M+1}^{M,L} \frac{u_i}{v_i} \frac{v_j}{u_j} P^{\dagger}_j P_i \right]^{l}  \left| \text{HF}\right\rangle ~.
\label{BCSP}
\end{equation}

It is important to note that this expansion in terms of particle and hole pairs contains the inverse of the factorial square. Having instead a simple factorial would lead to an exponential form that immediately connects the PBCS wave function to the pair coupled cluster theory of doubles (p-CCD). In fact, we have already explored the possibility of using these statistical coefficients as variational parameters \cite{Sierra2}. Moreover, in Ref. \cite{Hirsch} we compared PBCS with p-CCD and with the self-consistent RPA (SCRPA), showing that the latter approximations describe well the weak coupling limit, but fail dramatically when approaching the transition to superfluidity.
In order to proceed forward, we will derive the PBCS wave function in the ph basis starting from the
PBCS state expressed as a condensate of $M$ collective pairs, {\it viz.}

\begin{equation}
\left| \text{PBCS}\right\rangle =\frac{1}{\sqrt{\textsf{N}_{Cond}} }
\left[ \Gamma ^{\dagger }(x)\right] ^{M}\left|~ 0\right\rangle ~,\quad \Gamma
^{\dagger }(x)=\sum_{i=1}^{L}x_{i}P_{i}^{\dagger }  \label{pbcs} ~,
\end{equation}
where the norm $\textsf{N}_{Cond}$, which depends on the pair structure amplitudes $x_i$, can be obtained straightforwardly using the commutation relations (\ref{commu}) and recursive techniques \cite{Sierra2}.

In the PBCS approximation,  the $L$ structure coefficients $x_{i}$ of the condensed pair are considered as variational parameters chosen to minimize the expectation value of the hamiltonian. This approach is completely equivalent to the usual formulation of PBCS approximation \cite{PBCS}.

We now separate the collective pair operator $\Gamma^{\dagger}$ into its particle and hole components

\[
\Gamma ^{\dagger }(x)=\Gamma _{P}^{\dagger }(x)+\Gamma _{H}^{\dagger }(x)
\]
where
\begin{equation}
\Gamma _{P}^{\dagger }(x) =\sum_{p=M+1}^{L
}x_{p}P_{p}^{\dagger }~,~\Gamma _{H}^{\dagger } (x)
=\sum_{h=1}^{M}x_{h}P_{h}^{\dagger }  \label{ph}
\end{equation}

The PBCS state can then be rewritten as

\begin{eqnarray}
&&\left| \text{PBCS}\right\rangle =\frac{1}{\sqrt{\textsf{N}_{Cond} }}\left[ \Gamma _{P}^{\dagger }(x) +\Gamma
_{H}^{\dagger }(x) \right] ^{M}\left\vert 0\right\rangle = \nonumber \\
&&~=\frac{1}{\sqrt{\textsf{N}_{Cond} }} \sum_{l=0}^{M} \binom{M}{l} \left[ \Gamma _{P}^{\dagger }(x) \right] ^{l}\left[ \Gamma _{H}^{\dagger } (x) \right] ^{M-l}\left\vert 0\right\rangle ~. \nonumber \\
\label{phbcs}
\end{eqnarray}

The next step consists in replacing the vacuum state by the HF state of $M$ pairs. In doing so, we employ a trick that will be useful to rewrite the hole part of the PBCS condensate. Namely, we express the HF state as a condensate of $M$ collective pairs with the same amplitudes of the condensed PBCS pair, {\it viz.}

\begin{equation}
\left | \text{HF}\right\rangle =\frac{1}{\sqrt{\textsf{N}_{HF}}}\left[
\Gamma _{H}^{\dagger }(x)\right]^{M}\left\vert 0\right\rangle ~.
\label{HF}
\end{equation}
After some straightforward algebra, the PBCS wave function in the ph basis then reduces to

\begin{equation}
\left| \text{PBCS}\right\rangle =\sqrt{\frac{\textsf{N}_{HF} }{\textsf{N}_{Cond} }}\sum_{l=0}^{M}\frac{1}{l!^{2}}\left[ \Gamma
_{P}^{\dagger }\left( x\right) \Gamma _{H}\left( 1/x\right) \right]
^{l}\left| \text{HF}\right\rangle  \label{phbcs2}.
\end{equation}

This expression for the PBCS wave function is equivalent to Eq. (\ref{BCSP}),  displaying again the characteristic inverse of the factorial square. Moreover, the amplitudes of the hole destruction pair are the inverse of the amplitudes of the collective PBCS pair. By inserting the definition (\ref{ph}) for the particle and hole collective pairs, we obtain a more transparent form of the PBCS wave function

\begin{equation}
\left| \text{PBCS}\right\rangle =\sqrt{\frac{\textsf{N}_{HF} }{\textsf{N}_{Cond} }}\sum_{l=0}^{M}\frac{1}{l!^{2}}\left[ \sum_{p,h} \frac{x_p}{x_h} P_p^{\dagger } P_h
 \right]
^{l}\left| \text{HF}\right\rangle  \label{phbcs22}.
\end{equation}

As mentioned above the factorial square is the fingerprint of the PBCS wave function. Replacing these coefficients by a simple factorial allows us to define an Exponential form of the wave function,

\begin{equation}
\left| \text{Exp}\right\rangle = \frac{1}{\sqrt{\textsf{N}_{Exp}} } \sum_{l=0}^{M}\frac{1}{l!}\left[ \sum_{p,h} \frac{x_p}{x_h} P_p^{\dagger } P_h
 \right]
^{l}\left| \text{HF}\right\rangle  \label{phbcs3}~,
\end{equation}
which we will explore as an alternative to the PBCS approximation.

A third alternative approach, interpolates between PBCS and the Exponential form by using a new variational parameter $1\leq\alpha\leq 2$ in the exponent of the factorial, such that for $\alpha=1$ we recover the Exponential limit and for $\alpha=2$ we regain the PBCS state, {\it viz:}

\begin{equation}
\left| \text{Var}\right\rangle = \frac{1}{\sqrt{\textsf{N}_{Var}} } \sum_{l=0}^{M}\frac{1}{(l!)^{\alpha}}\left[ \sum_{p,h} \frac{x_p}{x_h} P_p^{\dagger } P_h
 \right]
^{l}\left| \text{HF}\right\rangle  \label{phbcsV}~.
\end{equation}

Coming back to the Exponential approximation (Exp)  in Eq. (\ref{phbcs3}), we note the similarity of this wave function with  p-CCD \cite{Hirsch, Scu1} and the particle-particle RPA  or SRPA \cite{Hirsch2} wave functions in the quasi-boson approximation, both of which have an exponential form,

\begin{equation}
\left| \Psi \right\rangle \propto \sum_{l=0}^{M}\frac{1}{l!}\left[ \sum_{p,h} X_{p,h} P_p^{\dagger } P_h
 \right]
^{l}\left| \text{HF}\right\rangle  \label{CCD},
\end{equation}
where the entries of the structure matrix $X_{p,h}$ are determined differently in Coupled Cluster and in RPA. Note that here too the statistical factor is the inverse of $l!$.

We should emphasize here two crucial differences between these two classes of approximation. In the first place, the approximations related to the PBCS wave function have a restricted separable form of the structure matrix $X_{p,h}=x_p/x_h$ which defines a single condensed pair and therefore takes into account two-body correlations only. p-CCD and SRPA preserve the complete freedom of the structure matrix, allowing for the inclusion of quartet correlations. The separable form of the structure matrix permits the explicit computation of norms and expectation values of operators and, therefore, the implementation of a Ritz variational theory. On the contrary, SCRPA and p-CCD need to resort to some approximations to evaluate expectation values, due to the non-separability of the structure matrix. In the case of RPA approaches this is done by means of a quasi-boson approximation, whereas p-CCD computes expectation values projectively in a subspace of zero and two ph state. As we will see in the numerical study to follow, the factorization of the matrix $X$ as in PBCS approaches leads to a poor approximation in the weak coupling limit dominated by pairing fluctuations.

\section{The exactly solvable BCS Hamiltonian as a benchmark model}

As noted in the previous section, we would like to test the ordinary PBCS and the associated Exponential and Variational approximations with the p-CCD approximation. Furthermore, we would like compare all of these methods with exact results where applicable. Since the Richardson method can be used to obtain exact solutions for the reduced BCS Hamiltonian, we have chosen to carry out the tests using such a hamiltonian acting in a set of doubly-degenerate single particle orbits with equal spacing, the so-called picket-fence model. We will first describe in a bit more detail the model and then in the next section report results of the comparisons for a specific size of the model space, as a function of the strength of the pairing force relative to the single-particle spacing.

The picket-fence model involves a Hamiltonian
\begin{equation}
H = \sum_{i=1}^{L} \epsilon_i \hat{N}_i - G \sum_{i,i'=1}^{L} P^{\dagger}_iP_{i'}~,
\end{equation}
with the equally-spaced single-particle energies given by
\begin{equation}
\epsilon_i - \epsilon_{i-1} = \epsilon~,
\end{equation}
whose exact eigenstates are given by the Richardson's ansatz
\begin{equation}
\left| \Phi \right\rangle \propto \prod_{\beta=1}^{M} \left[\sum_{i=1}^{L} \frac{1}{\epsilon_{i} -E_{\beta}} P_i^{\dagger} \right]\left| 0 \right\rangle ,
\label{Rich}
\end{equation}
where the $M$ pair energies $E_{\beta}$ are a particular solution of the set of $M$ non-linear coupled Richardson equations.
We will consider a system of $N=L$ particles, namely half filling, and discuss the results as a function of the pairing strength $G$ in units of the splitting between levels $\epsilon$.

\section{Test results of the PBCS and other approximate methods}

We present here benchmark results for a model with 20 double-degenerate levels at half filling. We show in figure 1 the relative error in the ground state correlation energies with respect to the exact Richardson solution  for the PBCS, Exponential, Variational and p-CCD  approximations.  The relative error is defined as $\Delta E= 1- E_c/E_c^{Exact}$. Results are reported in units of $\epsilon$ for values of the pairing strength ranging from weak coupling ($G < G_c$) to strong coupling ($G > G_c $), where the critical value $G_c$, determined by the solution of the BCS equations, separates the fluctuation-dominated regime with zero gap to the superconducting regime with finite gap.

\begin{figure}[htb]
\begin{center}
\includegraphics[height=.25\textheight]{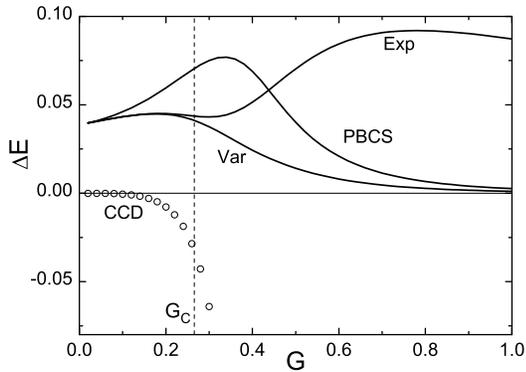}
\end{center}
\caption{Relative error with respect to the exact Richardson solution in the correlation energy calculated in the PBCS, Variational, Exponential and p-CCD approximations.  Results are reported for pairing strengths ranging from weak coupling ($G < G_c$) to strong coupling ($G > G_c $). Also shown is the critical value of the pairing strength $G_c$ in BCS approximation.}
\label{ener20}
\end{figure}

As can be seen in the figure, none of the approximations based on the PBCS wave function can describe appropriately the weak coupling limit. In particular, the three PBCS-like approximations do not approach smoothly the HF limit for $G\rightarrow 0$, despite the fact that the three approximations give the exact energy for $G=0$. This latter fact led to a wrong interpolation in the limit of $G\rightarrow 0$ in previous numerical studies of PBCS for this model \cite{Hirsch, Scu1}. In this limit, the dependence of the expansion coefficients on the factorial behavior is irrelevant, as shown by the three approximations converging to the same correlation energy. Contrary to the PBCS-like approximations, p-CCD correctly describes this limit due to the full freedom maintained in its structure matrix $X$, which is needed to reproduce the pairing fluctuations that dominate in this region. However, it progressively degrades as the system goes to the crossover region and fails completely in the superconducting phase up to the point at which there is no solution of the p-CCD equations. The figure also shows clearly the Ritz variational character of the approximations based on PBCS, which always give an upper limit to the GS energy. In contrast,  p-CCD severely overbinds for moderate values of $G$.
Within the PBCS class of approximations, the Exponential approximation describes better than PBCS the weak coupling region while PBCS is more efficient in the strong coupling region, producing the exact results in the large $G$ limit. As expected, the Variational approximation, which makes use of the extra degree of freedom provided by the parameter $\alpha$ to  interpolate between the two approximations, gives the optimal description along the complete crossover.
\begin{figure}[htb]
\begin{center}
\includegraphics[height=.25\textheight]{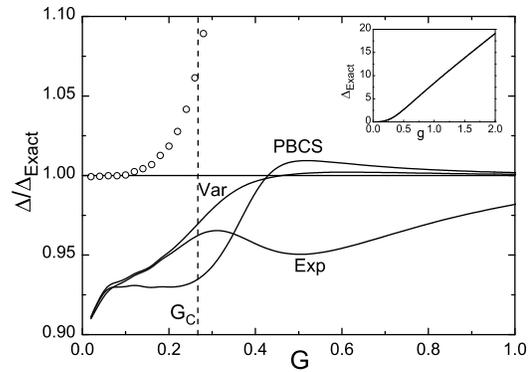}
\end{center}
\caption{Ratios of the canonical gaps to the exact Richardson gap calculated for the PBCS, Exponential, Variational and p-CCD approximations as a function of $G$.  We also show in the inset the behavior of the exact gap as a function of $G$.}
\label{gap}
\end{figure}

A similar pattern is observed in the canonical gaps, shown in figure \ref{gap} for the three PBCS-like approximations and for p-CCD. The canonical gaps are defined as $\Delta=G \sum_i \sqrt{ n_i (1-n_i)}$,  where the occupation probabilities $n_i$ are obtained in all cases using the Hellmann-Feynman theorem. The canonical gaps are more sensitive than correlation energies to the details of the wave function through their dependence on the occupation probabilities. As a consequence, the deviations from the exact gap are more pronounced than those of the correlation energies in the fluctuation-dominated regime. The Variational and PBCS approximations quickly converge to the exact result in the superconducting region. This is not the case for the Exponential approximation, which is not able to reproduce either correlation energies or gaps in the superconducting region. Similarly to the correlation energies, the p-CCD gap describes correctly the weak coupling limit, however the gap increases dramatically for moderate values of $G$ explaining the overbinding observed in Fig. \ref{ener20}.

\begin{figure}[htb]
\begin{center}
\includegraphics[height=.25\textheight]{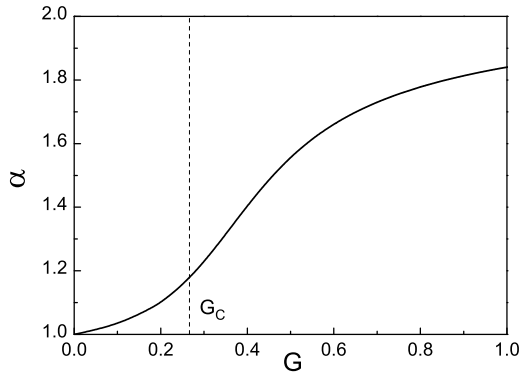}
\end{center}
\caption{The variational parameter $\alpha$ as a function of the pairing strength $G$.}
\label{Alfa}
\end{figure}

The variational parameter $\alpha$ defined in Eq. (\ref{phbcsV}) interpolates between the Exponential and the PBCS wave functions giving always the best correlation energy and gap as a function of $G$. The behavior of this parameter can be seen in Fig. \ref{Alfa}. As is evident from the figure, it starts with $\alpha =1$ for $G=0$, and increases smoothly across the transitional region towards the PBCS value $\alpha=2$, which is expected for the extreme superconducting limit.

\begin{figure}[htb]
\begin{center}
\includegraphics[height=.25\textheight]{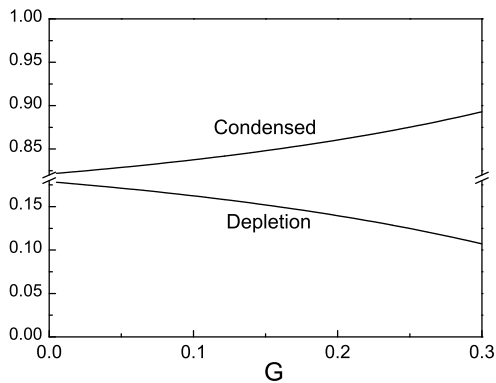}
\end{center}
\caption{The largest eigenvalue (Condensed) and the sum of all other eigenvalues (Depletion) of the structure matrix $X_ph$ in the p-CCD approximation as a function of $G$}.
\label{eigen}
\end{figure}

As mentioned above, the reason why the p-CCD approximation works well in the fluctuation dominated region is that its structure matrix $X_{ph}$ is non-separable. We can best analyze the structure of $X_{ph}$ in the p-CCD approximation by diagonalizing it. Due to the ph symmetry of the problem at half filling, $X_{ph}$ is a symmetric square matrix so that its diagonalization is equivalent to a singular value decomposition. Figure \ref{eigen} displays the largest eigenvalue and the sum of all other eigenvalues as a function of the pairing strength $G$. To facilitate the interpretation of the figure, we have normalized the trace of the matrix to 1. In analogy with a condensation phenomenon, we call the largest eigenvalue Condensed and the sum of the others Depletion. A separable matrix would have a unique eigenvalue different from 0. Therefore the Depletion is a signature of correlations beyond pairs. In particular, p-CCD takes into account quartet correlations which seem to be essential to describe correctly the fluctuation dominated regime. For increasing pairing strength, p-CCD collapses due to its rigid exponential structure.

\section{Summary and concluding remarks}

In this work, we have discussed several approximate methods for treating pairing. With atomic nuclei in mind, we have focussed on methods that exactly conserve the particle number. Several of the methods we considered are variational improvements to the PBCS approximation, deriving from a ph expansion of the PBCS wave function. The other method we considered is the non-variational p-CCD approximation.

Calculations were reported for a picket-fence model, whereby a pure pairing force acts in a space of doubly-degenerate equally-spaced levels. The calculations were carried out for a typical size nuclear system of involving 20 doubly-degenerate levels and 20 nucleons. It is worthwhile to mention here that the accuracy of PBCS increases with smaller size systems \cite{Sandulescu}. However, for these sizes it is always possible to carry out an exact diagonalization.

The model we studied exhibits a phase transition in mean-field BCS approximation at a critical value of the pairing strength $G_c$ relative to the splitting between levels.  In the superconducting region, both the PBCS and the Variational approximations converge to the exact results given by the Richardson solution. In the fluctuation dominated region, the Exponential and Variational approximations behave better than PBCS, but with a non negligible error relative to the exact solution.
The pair Coupled Cluster Doubles (p-CCD) approximation  is able to reproduce the exact results in the weak-coupling limit. This is because p-CCD exploits the complete freedom of the structure matrix, whereas the PBCS-related approximations use a restricted separable form of this matrix. While the Variational approximation produces an improvement over PBCS around the critical strength, it is actually worse than p-CCD  for strengths below the critical value. Indeed, in the weak-coupling limit it converges to exactly the same solution as both the PBCS and Exponential methods.

Further understanding of the behavior of these various approximate treatments of pairing correlations in the weak-coupling limit is clearly warranted. On the one hand, we showed that it is possible to extend variationally PBCS by using a new parameter $\alpha$ that interpolates between the exponential form, seemingly the preferred one in the weak coupling limit, to the PBCS form that provides exact results in the extreme superconducting limit. However, this variational theory is not able to capture completely the pairing fluctuations in the weak-coupling limit. The second ingredient required to fully describe pairing correlations across all regimes is the use of a non-separable structure matrix. The freedom of the structure matrix is preserved in p-CCD and particle-particle RPA, but at the cost of using a rigid exponential expansion and different approximate methods to evaluate norms and expectation values and thus breaking the Ritz variational principle. Moreover, the exponential form of both theories prevents accurate solution around the critical region. Possible frameworks that might be able to describe within the same formalism both phases are a variational theory that includes quartet correlations \cite{Sandu} or a non-variational theory like Coupled Cluster based on an expansion that interpolates between the Exponential and PBCS wave functions \cite{Tom}.

\begin{acknowledgements}
This work has been supported by the Spanish Ministry of Economy and Competitiveness under Grant No. FIS2012-34479, and in part by the US National Science Foundation under grant \# 0854873.

\end{acknowledgements}


\end{document}